\documentclass[]{spie}  
\usepackage{epsfig}
\usepackage{epstopdf}
\usepackage{yfonts, graphicx, wrapfig}
\usepackage{amssymb}
\usepackage{amsmath}
\usepackage{varwidth}
\usepackage{algorithm}
\usepackage{algpseudocode} 
\algnewcommand{\Inputs}[1]{%
  \State \textbf{Inputs:}
  \Statex \hspace*{\algorithmicindent}\parbox[t]{.8\linewidth}{\raggedright #1}
}
\algnewcommand{\Initialize}[1]{%
  \State \textbf{Initialize:}
  \Statex \hspace*{\algorithmicindent}\parbox[t]{.8\linewidth}{\raggedright #1}
}

\usepackage[backref=page]{hyperref} 
\hypersetup{
   pdfborder= 0 0 0
   citecolor = red,
   linkcolor = red,
}
\title{Anomaly detection in clutter using spectrally enhanced Ladar} 

\author{Puneet S. Chhabra\supit{a},  Andrew M. Wallace\supit{a} and James R. Hopgood\supit{b}
\skiplinehalf
\supit{a}School of Engineering and Physical Sciences, Heriot-Watt University, Edinburgh, UK \\
\supit{b}School of Engineering, University of Edinburgh, Edinburgh, UK}

\authorinfo{Further author information: (Send correspondence to Puneet S. Chhabra.)\\P.C: E-mail: psc31@hw.ac.uk, Telephone: +44 131 451 4168 }

\begin{document} 
  \maketitle 

\begin{abstract}
Discrete return (DR) Laser Detection and Ranging (Ladar) systems provide a series of echoes that reflect from objects in a scene. These can be first, last or multi-echo returns. In contrast, Full-Waveform (FW)-Ladar systems measure the intensity of light reflected from objects continuously over a period of time. In a camouflaged scenario, e.g., objects hidden behind dense foliage, a FW-Ladar penetrates such foliage and returns a sequence of echoes including buried faint echoes. The aim  of this paper is to learn local-patterns of co-occurring echoes characterised by their measured spectra. A deviation from such patterns defines an abnormal event in a forest/tree depth profile. As far as the authors know, neither DR or FW-Ladar, along with several spectral measurements, has not been applied to anomaly detection. This work presents an algorithm that allows detection of spectral and temporal anomalies in FW-Multi Spectral Ladar (FW-MSL) data samples. An anomaly is defined as a full waveform temporal and spectral signature that does not conform to a prior expectation, represented using a learnt subspace (dictionary) and set of coefficients that capture co-occurring local-patterns using an overlapping temporal window. A modified optimization scheme is proposed for subspace learning based on stochastic approximations. The objective function is augmented with a discriminative term that represents the subspace's separability properties and supports anomaly characterisation. The algorithm detects several man-made objects and anomalous spectra hidden in a dense clutter of vegetation and also allows tree species classification.
\end{abstract}



\section{INTRODUCTION} \label{sec:intro} 
\textbf{La}ser \textbf{d}etection \textbf{a}nd \textbf{r}anging (Ladar) is an active depth (spatial) sensing modality, operated from ground, aerial and space-based platforms. This has found a variety of applications in remote sensing and military scenarios for its ability to extract spatial information from a 3D scene. Passive Multispectral (MSI) and Hyperspectral Imaging (HSI) with dozens and hundreds of wavelengths respectively, measure spectral signatures for each individual image pixel with no spatial information. A fusion of spectral imagery with 3D spatial information may seem desirable. This comes at a cost of both processing time and complexity. The evolution from passive MSI and HSI into spectrally aware active Laser sensors seems natural since simultaneous measurements can eliminate data synchronization errors (both in space and time). Full-waveform Multispectral Ladar (FW-MSL) signals are active spatial measurements that are made simultaneously at different wavelengths using Ladar based sensors. In the past, Ladar sensing has seen applications in ecological studies for tree species classification and segmentation \cite{reitberger2006}, geo-morphology \cite{brodu2012} and foliage filtering \cite{jutzi2006}. The relevance of urban scene classification with applications to remote sensing has also been studied extensively. \cite{jutzi2006,mallet2011,chehata2009}. 

More recently, small foot-print sensing using MSL has been possible in retrieving structural and physiological properties of vegetation\cite{wallace2010}. Although popular with the remote sensing community, the choice of MSL sensors and raw waveforms for anomaly detection in complex environments (urban city or dense forest) has not been investigated. Echo characterisation and classification is an active area of research but this requires sophisticated echo modelling algorithms which operate on the entire dataset. The aim here is to propose a framework that is smarter in its search for anomalies. 

Automatic Target Recognition (ATR) are scenarios that may benefit, for example, in automatic detection of known targets hidden or camouflaged under dense foliage. The nature of anomaly detection differs from the ATR notion. The type of target, its structure and material may be unknown. Consider a surveying scenario where a forest is scanned looking for normality and abnormality in what is being perceived through different sensor modalities. A set of measurements can be made on what is being observed and one may be interested in narrowing down the target search window by finding spatial and spectral anomalies for further structural analysis. This can be carried out in two phases:
 \begin{figure}[!t]
  \centering
 \includegraphics[width=17.3cm, height=6cm]{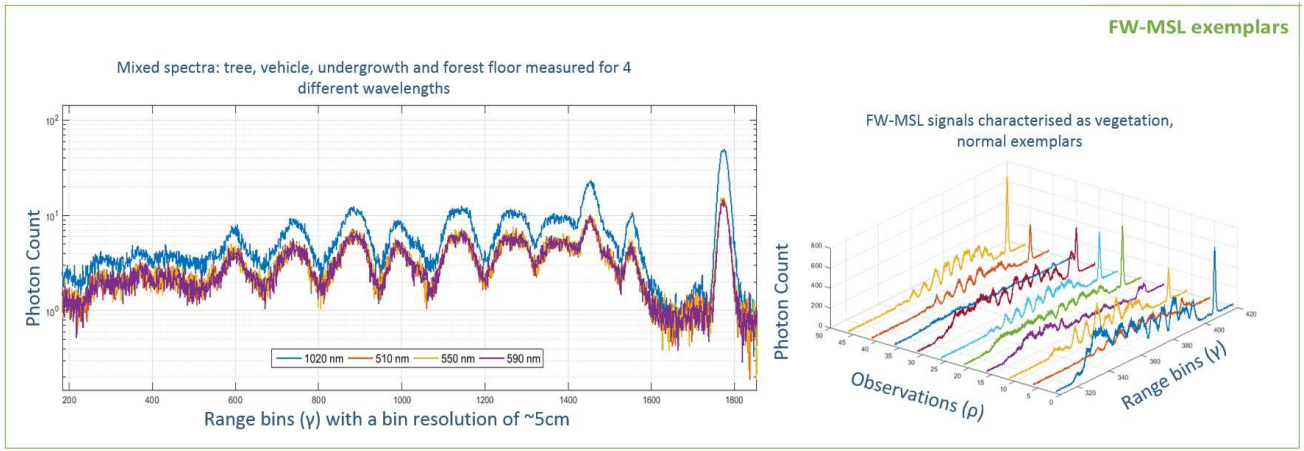}
  \caption{Example FW-MSL signals used to detect anomalies. (Left) A typical FW-MSL return of a large foot-print with a range resolution of $\sim5cm$ per bin. The backscattered returns shown here are recorded for 1020, 510, 550 and 590 nm wavelengths. This depth profile is a mixture of different spectra: vehicle, conifer, undergrowth and bare-ground. When observed from an altitude the tree top is the first peak, $600^{th}$ bin on the left. (Right) Stacked FW-MSL exemplars. }
  \label{fig:example}
 \end{figure}
\begin{itemize}
  \item \textit{Phase 1} of the survey operates on a \textit{large-footprint} setting which aims to identify regions with spectral and spatial (time) anomalies, i.e., similar spectral signatures found at depths which do not comply with learnt models. This paper proposes a method to support Phase 1. Once these regions are identified, Phase 2 is applied to regions that contain such anomalies. 
  \item \textit{Phase 2} A \textit{small-footprint} scanning scheme can be employed along with expensive signal modelling algorithms to deliver a dense 3D point cloud for structural analysis.
\end{itemize} 

In order to support Phase 1 and Phase 2, Ladar systems have been designed and developed operating on simultaneously upto 4 wavelengths \cite{wallace2014}. In order to support future instrument design and parameter selection experiments on simulated and synthetic data have been carried out on a range of wavelengths: $4$, $8$, $16$ and $32$. Figure 1 illustrates example FW-Ladar signals measured over a plot of vegetation. 

\subsection{Contributions} \label{contri}
In this paper, a two-stage surveillance framework involving a sensing scheme that makes use of both small and large footprint Ladar scanning is proposed. In order to model natural trees or forest plots, a layered representation is considered that conforms to a sparse approximation and discrimination model. The framework includes: 
\begin{itemize}
  \item This work aims to avoid computationally expensive signal modelling algorithms in order to characterize backscattering echoes for target localisation on large FW-MSL datasets. To support such an aim, a cueing algorithm that reduces target search space by identifying anomalies in a large urban or forest scene is proposed. Such an approach is novel since it is can be an efficient way of combining large and small foot-print scanning, especially when surveying large environments and the multi-spectral nature of the sensors used.
  \item A semi-supervised subspace (a set of linearly dependent basis, i.e., dictionary) learning method (Section \ref{sec:approach}) which learns normal Ladar signals of different conifer species and forest terrain comprising of 8 different spectra (bark, needle, terrain, under-growth, steel, concrete, brick and clay tiles); and each new observation is transformed and reconstructed using the learnt dictionary. High reconstruction errors reveal hidden anomalies. The optimisation routine used here is influenced from the compressed sensing literature except that objective function is modified in order to classify tree species and other signals. See Algorithm~\ref{alg:sad}. 
  \item The proposed framework can emphasize any unknown as opposed to a known signature \cite{harsanyi1994hyperspectral} both in spatial and spectral context. To the best of authors knowledge, this paper is the first attempt at applying such techniques to foliage penetration and anomaly detection on raw FW-MSL measurements. 
  \item This work also shows that the proposed approach is widely applicable as the definition of an anomaly is not restrictive to a particular target type. The concept of context is associated with different FW-MSL measurements across different layers. In the past, this has been over-looked, echo modelling schemes have treated each individual echo independently and characterise with its location and amplitude. This work shows that relationship learning across layers and wavelengths can discover co-occurring patterns which can be highly relevant to anomaly detection.
\end{itemize}

\subsection{Outline} \label{outline}
An outline of the paper is as follows: Section \ref{sec:problem} provides a mathematical formulation of the problem and description of an anomaly. Section \ref{sec:approach} offers a solution to the problem with experiments and results on the data in Section \ref{sec:Experiments}. Finally, Section \ref{sec:Conclusion} provides conclusions and future work.
\begin{figure*}[!t]
\centering
 \includegraphics[width=17.3cm, height=4.8cm]{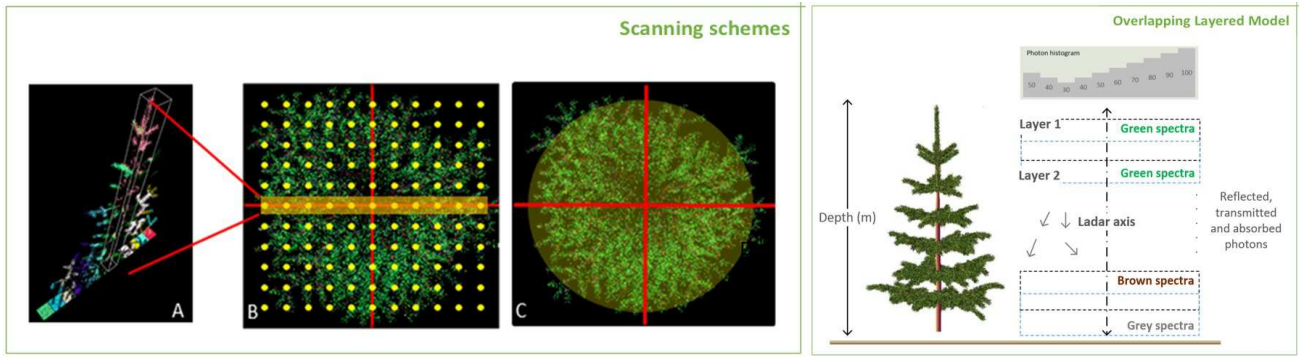}
  \caption{(A) A cross-section of a single conifer tree. (B) A \textit{push-broom} style (small-foot print) dense scan pattern resulting in a cross-section slice of a tree. A \textit{flash} style (large-foot print) scan pattern is illustrated in (C). (D) The forest/tree layered model.} 
  \label{fig:scan_schemes}
 \end{figure*} 
\section{\uppercase{ \textbf{Problem Definition }}} \label{sec:problem}
Anomaly detection can be defined as a problem of detecting data samples that significantly deviate from an expected model. Methods that can capture essential patterns both in temporal and spectral dimensions within the original data space are desirable. Such patterns contribute significantly to the generating cause of an observation, a measured signal, and such representation are in a lower dimensional space.

Recent developments in \textit{signal approximation} \cite{candes2008introduction} and dimensionality reduction have an underlying concept, \textit{Sparsity}, which means that an original signal, which is dense in certain basis, once its is transformed in to another convenient basis i.e., the contribution vector is almost sparse. In this paper, it is assumed that an expected prior conforms to a sparse representation model, where the coefficients represent the contribution of a set of \textit{atoms}, together called a subspace. This work also employs method that extract and learns such a subspace and their coefficients from raw FW-MSL data. Together, they contribute towards an \textit{accurate} representation of the original FW-MSL measurement data. The following section introduces a layered representation of a forest, followed by an FW-MSL approximation model.
  
\subsection{Layered representation of forest/tree} \label{sec:layered}
	An explicit assumption is made here that a depth profile through a dense forest or an individual tree can be represented by a series of instrumental Ladar returns from a set of \textit{layers} (see Figure \ref{fig:scan_schemes}D) at different ranges. Thus, layer position defines the \textit{context} of a tree and the number of layers, the number of individual spectra measured within a layer, and their contribution defines its \textit{behaviour}.
	  
	A depth profile of a tree or forest are made up of countless surface responses from leaf, bark and other man-made surfaces found at varying depths. This is a \textit{leaf-on} condition, dense healthy forests, mostly during summer. During winter, a \textit{leaf-off} condition, most of the laser light penetrates trees quiet easily. Such a condition is over-looked at this stage. The illustration in Figure \ref{fig:scan_schemes} highlights one such example with spectral response registered at each layer. The top-most layer records green, since it mostly consists of high chlorophyll content representing natural material. The bottom-most layer records brown dead leaves and undergrowth and finally forest floor records spectra different to other layers, color coded here as grey. This sequence clearly deviates under diverse vegetation or with varying tree heights. Hence, the intuition here is to capture the relationship between over-lapping layers in order to capture any co-occurring patterns.
	
	This can be achieved by breaking down a data sample, in its entirety in to several neighbouring sub-sequences and measure the co-occurrence of local-patterns within every sub-sequence. For the sake of simplicity, running an overlapping sliding window over wavelength$\left(\lambda\right)$-range$\left(\gamma\right)$ series resulting in multiple sub-sequences. The aim is to learn lower-dimensional subspaces in order to seek unusual local patterns that do not appear in reference range-amplitude series. An implicit assumption made here is that the flying altitude is known and that each individual FW-MSL measurement is registered with respect to the flying altitude, i.e., that the length of the two samples and the range bin index are same. Next, the concept of normality and abnormality in the context of range-amplitude-wavelength, FW-MSL data, is explained.
	
\subsection{The concept of normality and abnormality} \label{sec:anomaly_space}
 In a semi-supervised anomaly detection technique, it is assumed that a database of \textit{normal} exemplars that conform to non-anomalous data is known. Within the realm of this work, these exemplars may belong to several different classes, i.e., different tree species or Ladar returns of vegetation and bare forest floor. A continuous range-amplitude depth profile of a tree or forest is made up of several different spectra: bark, needle, dead leaves, undergrowth and forest floor. Such data samples are qualified as \textit{normal}, free from any man-made objects. Deviation from such learnt model qualifies as \textit{abnormality}.

Scenes observed using active/passive range sensors are made up of objects of interests, background clutter, sensor and environmental noise. Background clutter can be defined as something that is neither anomalous nor an object of interest. Interestingly, an anomaly characterised by a context or an outcome of a condition can be referred as a \textit{contextual anomaly}, or a \textit{conditional anomaly}. In the context of FW-Ladar measurements, the input data is a range-amplitude-wavelength series and the context here is occurrence of an event at a particular range $\gamma$ or wavelength $\lambda$.

 In a layered representation (Figure \ref{fig:scan_schemes}) an anomaly can occur in any range bin, i.e., depth. This could be due to the variability in tree height across a section of the forest, tree health (climatic conditions) or a more interesting cause: presence of man-made object(s). It is highly impractical to collect data samples for all abnormal cases, but, what is available in abundance is large amount of training data, normal events only, plots of natural forests with trees. The anomaly detection problem can be defined as follows: 
 \begin{quote}
 	\textit{One wishes to learn the abundance of distinctive material spectra and their relationships across different layers within a range-amplitude series. The aim is to detect a deviation from the learnt \textit{normal} models resulting in an abnormal event.}
 \end{quote} 

\subsection{Waveform Model}
A set of Ladar signals measured for an individual wavelength $\lambda$ can be defined as:
\begin{equation}
	\mathit{\textit{L}}_{\lambda} = \left\{ l_{\lambda,1},...,l_{\lambda,\Gamma}\right\},
\end{equation} where $\Gamma$ is the number of of range bins, i.e., $\gamma \in \left[1,2,...,\Gamma\right]$ and $\lambda \in \left[1,2,...,\Lambda\right]$. Formally, $l_{\lambda,\gamma}$ expresses the photon count within the $\gamma^{th}$ channel (range bin) of the $\lambda^{th}$ spectral band being observed. Here, the assumption is that each photon count $l_{\lambda,\gamma}$ is drawn independently from a Poisson distribution: 
\begin{equation} \label{eq:poisson_distribution}
		Prob(L_{\lambda})=e^{-\delta_{\lambda,\gamma}}\,\frac{\left(\delta_{\lambda,\gamma}\right)}{l_{\lambda,\gamma}!}^{\, l_{\lambda,\gamma}}, 
\end{equation}
In \ref{eq:poisson_distribution}, $\delta_{\lambda,\gamma}$ is defined as: 
\begin{equation}
	\delta_{\lambda,\gamma} = s_{\lambda,:}K_{0,\lambda}(\gamma) + b_{\lambda}, 
\end{equation}
where $s_{\lambda,:}$, denotes different surfaces a reference beam interacts with, $K_{0,\lambda}(\cdot)$ is the photon impulse response, and $b_{\lambda}$ is background and dark photon level across all bins. The aim of this work is not to model or deconvolve such responses in order to find echoes. Instead, the proposed approach aims to operate on raw FW-MSL signals collected over a synthetic forest in order to leave sophisticated echo modelling algorithms to operate on only a sub-set of \textit{interesting} regions in the scene.  

\subsection{Mapping raw Ladar data into sub-spaces} \label{sub_learning}
FW-MSL signals measured for $\Lambda$ wavelengths can be expressed as:
\begin{equation}
\mathbf{L} =  \left[\mathit{L_{1}},...,\mathit{L_{\Lambda}}\right]^{T}.
\end{equation}
The original range-amplitude-wavelength series $\mathbf{L}$ for different wavelengths $\left(\Lambda\right)$ is mapped on to a set of overlapping sub-sequences $\mathbf{W}$ and then transformed in to a feature subspace $\mathbf{F}$. The mapping can be written as follows: 
 	\begin{equation} \label{l_to_w}
 		\mathbf{L} \to \mathbf{W} \to \mathbf{F}
 	\end{equation} 
The first half of the mapping is explained below followed by the final mapping into $\mathbf{F}$ is explained in more detail at the end of Section \ref{sec:learning_discrimin_sub} using \ref{F_matrix} and \ref{eq:W_to_F_final}. 
\begin{figure}[!t]
\centering
 \includegraphics[width=17.3cm, height=5.8cm]{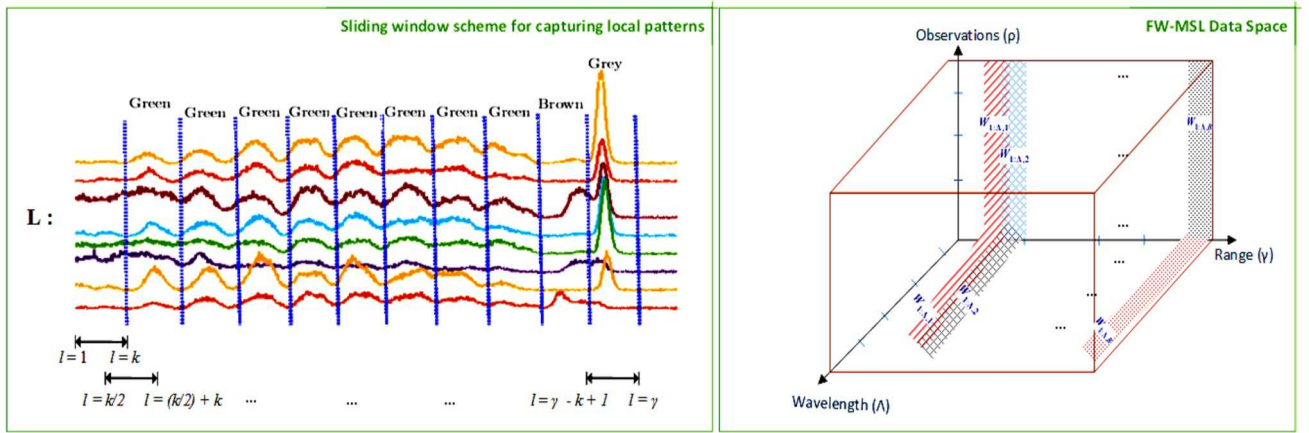}
  \caption{Sliding window concept in order to extract over-lapping range-amplitude series.} 
  \label{fig:sliding_concept}
 \end{figure} 
\paragraph{Overlapping sub-sequence representation ($\mathbf{L}\to\mathbf{W}$)} Figure \ref{fig:sliding_concept} illustrates the windowing concept and subspace representation of normal vs abnormal events occurring in the range-amplitude-wavelengths space. $\mathbf{L} \in \mathbb{R}^{\Gamma\times\rho\times\Lambda}$ for $\rho$ training samples, $\Lambda$ number of wavelengths and $\gamma$ number of range bins. So the $\mathbf{L}\to\mathbf{W}$  can be written as:
\begin{equation} \label{eq:L_to_W}
	\mathbf{W} = \left[W_{\lambda}, ... ,W_{\Lambda}\right]^{T}, 
\end{equation} where $W_{\lambda}$ is a sub-sequence representing a layer of the forest/tree of size $B=\frac{k}{\Gamma}$ that are partitioned on the range ($\gamma$-axis), where $k$ controls the window size and $m$ is the amount of overlap. For a fixed wavelength $\lambda$ this can be written as: 
\begin{equation} \label{eq:L_to_W_final}
W_{\lambda}=\left\{ \underset{W_{\lambda,1}}{\underbrace{l_{\lambda,1}\,,\,...\,,\, l_{\lambda,k}}}\,\,,\,\,\underset{W_{\lambda,2}}{\underbrace{l_{\lambda,1+m}\,,\,...\,,\, l_{\lambda,k+m}}}\,\,,\,\,...\,\,,\,\,...\,\,,\underset{W_{\Lambda, B}}{\underbrace{l_{\Lambda,\Gamma-k+1}\,,\,...\,,\, l_{\Lambda,\Gamma}}}\right\}
\end{equation}

\paragraph{Subspace representation} Each data sub-sequence, $W_{\lambda, i}$, where $i=\left[1,2,...,B\right]$ is the sequence index in \ref{eq:L_to_W_final} can be described with an over-complete basis (dictionary or subspace), $D_{\lambda, i}$, a coefficient matrix $C_{\lambda, i}$ plus additive constant background noise:
\begin{equation} \label{eq:sparse_model}
	W_{\lambda, i} \approx D_{\lambda, i} C_{\lambda, i},
\end{equation}
where $W_{\lambda, i} \in \mathbb{R}^{k \times \rho }$ and $D_{\lambda, i} \in \mathbb{R}^{k \times \eta }$ is an over-complete $\left( \eta \gg k \right)$ matrix and the resulting coefficient matrix $C_{\lambda, i} \in \mathbb{R}^{k \times \eta}$. The goal here is to find an optimal subset, called \textit{Subspace} (\textit{Dictionary}) $\left(D_{\lambda, i}\right)^{'}$ such that the set $W_{\lambda, i}$ can be accurately reconstructed by $\left(D_{\lambda, i}\right)^{'}$ and the size of this subspace is as small as possible. A constant background Poisson noise model in $W_{\lambda, i}$ is assumed. A simple way of doing this is to randomly and uniformly select basis vectors to build the subspace. This strategy can be risky as it may not capture the variability in the training data. A principled way of selection which selects an optimal subset of $D_{\lambda, i}$ as a subspace is presented below. 

Using all the vectors of $D_{\lambda, i}$, denoted by $\{d_{j}\}_{j=1}^\eta$ (see Section \ref{sec:learning_discrimin_sub}), all vectors of $C_{\lambda, i}$, denoted by $\{C_{\lambda, i}\}_{j=1}^{\eta}$ a signal $W_{\lambda, i}$ can be represented as a sparse linear combination of these columns, referred to as \textit{atoms}. The representation of $W_{\lambda, i}$ may either be exact (\ref{eq:sparse_model}) or an approximation $ ||W_{\lambda, i} - D_{\lambda, i}C_{\lambda, i}||_{2}\le\epsilon$. For a full-rank matrix $D_{\lambda, i}$ and $k<\eta $, there are an infinite number of solutions to this representation (\ref{eq:sparse_model}). A stop constraint is needed on this solution, either set a sparsest approximation constraint, i.e. $min_{C_{\lambda, i}} ||C_{\lambda, i}||_{0}$, where $||\centerdot||_{0}$ is the $\mathit{l}^{0}$ quasi-norm, which counts the non-zero entries of a vector or stop when the error term $\epsilon$ stops decreasing and goes below a threshold. An optimal can be defined by minimising the following objective function:  
\begin{equation} \label{eq:sparse_solution}
	arg\; \underset {D_{\lambda, i},C_{\lambda, i}}{min}\left[||W_{\lambda, i} - D_{\lambda, i}C_{\lambda, i}||_{2}^{2} + \beta_{1}\,||C_{\lambda, i}||_{0} \right], 
\end{equation}
where the parameter $\beta_{1} > 0$ is a regularization scalar that balances the trade-off between the reconstruction accuracy and sparsity.  Finding a solution to (\ref{eq:sparse_solution}) is an NP-hard problem primarily due to its combinatorial optimisation, i.e., minimising for best reconstruction with least number of basis vectors are convex individually but not convex jointly. This can be solved iteratively.  

A suboptimal solution to this problem is found by replacing the $l_{0}$ norm in (\ref{eq:sparse_solution}) with the $l_{1}$ norm and re-write (\ref{eq:sparse_solution}) using components of $C_{\lambda, i}$ as: 
\begin{equation} \label{eq:sparse_solution_l1}
	 arg\; \underset {D_{\lambda, i},C_{\lambda, i}}{min}\left[||W_{\lambda, i} - D_{\lambda, i}C_{\lambda, i}||_{2}^{2}  + \beta_{1}\, \sum_{j = 1}^{\eta} ||\{C_{\lambda, i}\}_{j}||_{1} \right], 
\end{equation}
where $||\centerdot||_{1}$ is the $l_{1}$ norm. In the following section an iterative way to solve \ref{eq:sparse_solution_l1} is explained in more detail.
\renewcommand{\algorithmicrequire}{\textbf{Input:}}
\renewcommand{\algorithmicensure}{\textbf{Output:}}
\begin{algorithm}[!tp]
\caption{\textbf{Anomaly Training}}
\label{alg:training}
\begin{algorithmic}[1]
\Require{Configuration file with $\Gamma, \rho, \Lambda, k, m, weights, mode$ and $\eta$ }
\Ensure{Learnt dictionaries and coefficients, i.e., $\mathbf{L}\to \mathbf{W}\to\mathbf{F}$}
\Procedure{anomalyTrain}{configFile}
	\State \textbf{normalise} $\mathbf{L}$, FW-MSL data
	\For{i = 1 to num\_of\_Layers}
		\State \textbf{Map} $\mathbf{L} \to \mathbf{W}$ using \ref{eq:L_to_W_final} 
		\State \textbf{Call: }$[\,\mathbf{C}_{\lambda,i}$, optimized $\mathbf{D}_{\lambda,i}\,]${ = SAD}({$\mathbf{W}_{\lambda,i}$, $\mathbf{D}_{\lambda,i}$, mode, weights}) - Algorithm~\ref{alg:sad} \Comment see Section \ref{sec:learning_discrimin_sub}  
	     \State \textbf{Learn local-patterns} on joint-wavelength coefficient matrices  \Comment See section \ref{sec:rln}
	     \State \textbf{Transform} layer wise coefficients of $\mathbf{W}_{\lambda,i}$ to a lower dimensional space \Comment See section \ref{sec:rln}
	     \State \textbf{Map} $\mathbf{W}\to \mathbf{F}$ \Comment Equation \ref{F_matrix} and\ref{eq:W_to_F_final}
	     \State \textbf{Save} $\mathbf{F}_{train}$
	\EndFor
\EndProcedure
\end{algorithmic}
\end{algorithm}
\section{\uppercase{Adopted Approach}} \label{sec:approach}
	
A solution to (\ref{eq:sparse_solution_l1}) leads to a good approximation of measurement matrix $W_{\lambda, i}$. A discriminating trait is added to the original problem and finally the reconstruction error is used to differentiate normal from abnormal data. This framework also allows for classifying of test ladar signals in to different classes (e.g., tree species). A natural forest environment is made up of several tree species with variable tree height. Hence, one cannot fix one tree \textit{layer} model for different tree species. One solution to capture such variability is to have species bound subspaces and coefficients. 

As illustrated in the previous section the data is layered in to overlapping sub-sequences. The training step is shown in the form a pseudo code in Algorithm~\ref{alg:training}. The optimisation routine with an discriminative objective function is presented in Algorithm~\ref{alg:sad}. For the sake of completeness the optimisation routine is presented in Algorithm~\ref{alg:sad}. The complete approach can be briefly listed in three steps: 
\begin{enumerate}
	\item \textit{Subspace selecting \& Learning} In training mode a matching pursuit method is employed in order to learn optimal subspaces, $(D_{\lambda, i})^{'}$ and multi-wavelengths coefficient matrix $\mathbf{F}$, i.e., $\mathbf{W}\to\mathbf{F}_{train}$. Such a representation can be extended by including a class discriminative (non-anomalous) criterion in to the objective function such that during the approximation stage discriminating subspaces can be learned for different normal classes (conifer species, bare forest floor, and undergrowth). In testing mode, using the learnt subspace a new coefficient matrix $\mathbf{F}_{test}$ is computed.
	\item \textit{Relationship learning \& anomaly detection} $\mathbf{F}_{train}$ is decomposed using a semantic learning technique, transformed and reconstructed. A test sample is projected back to the subspace and reconstructed, a high reconstruction error reveals the anomalies present in the test data samples.  
\end{enumerate}


\subsection{Step 1: Learning discriminating subspaces and Coefficients} \label{sec:learning_discrimin_sub}
 When solving (\ref{eq:sparse_solution_l1}) The aim here is to approximate a subspace that allowsan assumption is made that the training data are labelled exemplars for only the normal class. Equation (\ref{eq:sparse_solution_l1}) can be updated with a discrimination function 
parametrised by $D_{\lambda, i}$ as a Fisher Discriminant which has been very popular with Linear Discriminant Analysis (LDA), which is a generalisation of Fisher's linear discriminant \cite{scholkopft1999fisher}. The discrimination power is maximised when the spatial distribution of samples from similar classes (scatter matrix, $S_{w}$) are closer to each other when compared to samples from different classes (scatter matrix, $S_{b}$). 
\renewcommand{\algorithmicrequire}{\textbf{Input:}}
\renewcommand{\algorithmicensure}{\textbf{Output:}}
\begin{algorithm}[!tp]
\caption{Simultaneous Approximation \& Discrimination (SAD)}
\label{alg:sad}
\begin{algorithmic}[1]
\Require{Measurement vector $W_{\lambda, i} \in \mathbb{R}^{k \times \rho}$; number of atoms, $\eta$; $weights=\left[\beta_{1},\beta_{2}\right]$, and mode}
\Ensure{Learnt dictionary $\left(D_{\lambda, i}\right)^{'}$; Coefficients $C_{\lambda, i}$}
\Procedure{SAD}{({$\mathbf{W}_{\lambda,i}$, $\mathbf{D}_{\lambda,i}$, mode, weights})}
	\State \textbf{Initialize:} Index set $\omega = \phi$ and residual $(R_{\lambda, i})_{0} = W_{\lambda, i}$
	\While{$R_{\lambda, i} \neq 0$}
		\State \textbf{Identify:}\quad An atom from subspace $D_{\lambda, i}$ that maximizes: 
		\If{$mode==approximate\_only$}
			\State \quad\quad\quad\quad $d: arg\; \underset {d\in D_{\lambda, i}}{max} J\left(d^{T}(R_{\lambda, i})_{\eta},\beta_{1}\right)$ \Comment Equation\ref{eq:sparse_solution_l1}
		\Else
			\State \quad\quad\quad\quad $d: arg\; \underset {d\in D_{\lambda, i}}{max} J\left(d^{T}(R_{\lambda, i})_{\eta},\beta_{1},\beta_{2}\right)$ \Comment Equation \ref{eq:fisher_objective}
		\EndIf
		\State \textbf{Compute: }Orthogonal projection $(P_{\lambda, i})_{\eta}$ (for remaining atoms):
		\State \quad\quad\quad\quad\quad\quad\quad $(P_{\lambda, i})\in \{d\}_{j=1}^{\eta -1} = D_{\lambda, i} \ast inv\left( (D_{\lambda, i})^{T} \ast D_{\lambda, i} \right) \ast (D_{\lambda, i})^{T}$
		\State \textbf{Compute: }The new residue is given as:  
		\State \quad\quad\quad\quad\quad\quad\quad $(R_{\lambda, i})_{\eta} = W_{\lambda, i} - (P_{\lambda, i})_{\eta}\,W_{\lambda, i}$
		\State \textbf{Increment: }$\eta = \eta + 1$
	\EndWhile
\EndProcedure
\end{algorithmic}
\end{algorithm}
For a \textit{subspace}, $D_{\lambda, i} = \left[d_{1},d_{2},...,d_{\eta}\right]$ (where $d_{1}, d_{2},...,d_{n}$ are the \textit{atoms} in the subspace, \ref{eq:L_to_W}), of which $\eta_{i}$ samples are in class $\varOmega_{i}$, for $1 \leq i \leq \varOmega$, mean is defined as $\mu_{i}$ and variance $\upsilon_{i}^{2}$ for class $\varOmega_{i}$ as $\mu_{i} = \frac{1}{\eta_{i}}\, \sum_{d \in \varOmega_{i}} \mathbf{d},$ and $	\upsilon_{i}^2 = \frac{1}{\eta_{i}} \sum_{d \in \varOmega_{i}} ||\mathbf{d}-\mu_{i}||_{2}^{2}$. 
The mean of all samples can be written as: $\mu = \frac{1}{\eta} \sum_{i=1}^{\eta}d_{i}$. Finally, the Fisher discrimination function for the proposed objective function is defined as: 
\begin{equation} \label{eq:fisher_dis}
	K\left(D_{\lambda, i}\right) =  S_{w}^{-1} S_{b},  
 \end{equation} where $S_{w}$, the \textit{inner-class} scatter matrix and $S_{b}$, the \textit{intra-class} scatter matrix are defined as follows: 
\begin{equation} \label{eq:fisher_dis}
S_{b} = ||\sum\limits_{i=1}^{\varOmega}\eta_{i}\left(\mu_{i}-\mu\right)\,\left(\mu_{i}-\mu\right)^{T}||_{2}^{2} \quad , \quad S_{w} = \sum\limits^{\varOmega}_{i=1}\upsilon_{i}^{2}
\end{equation}
Using (\ref{eq:fisher_dis}) update (\ref{eq:sparse_solution_l1}) with: 
\begin{equation} \label{eq:fisher_objective}
	arg\; \underset {D_{\lambda, i},C_{\lambda, i}}{min}\left[K\left(D_{\lambda, i}\right) + \beta_{1} \sum\limits_{j=1}^\eta||\{C_{\lambda, i}\}_{j}||_{1} + \beta_{2}\sum\limits_{j=1}^{\eta}||W_{\lambda, i} - D_{\lambda, i}\{C_{\lambda, i}\}_{j}||_{2}^{2}\right], 
\end{equation} 
where $\beta{1}$ and $\beta_{2}$ are positive scalar chosen as a trade-off between reconstruction error, coefficients contribution and discrimination. An iterative optimisation routine motivated by pursuit algorithms \cite{tropp2006algorithms,pati1993orthogonal} is suggested in order to solve \ref{eq:sparse_solution_l1}. It is coined as \textit{Simultaneous Approximation \& Discrimination (SAD)} in this work as a discriminating term is added to the original objective function.\footnote{If the classification mode is selected replace (\ref{eq:sparse_solution_l1}) by (\ref{eq:fisher_objective})}

\begin{figure*}[!t]
\centering
 \includegraphics[scale=0.4]{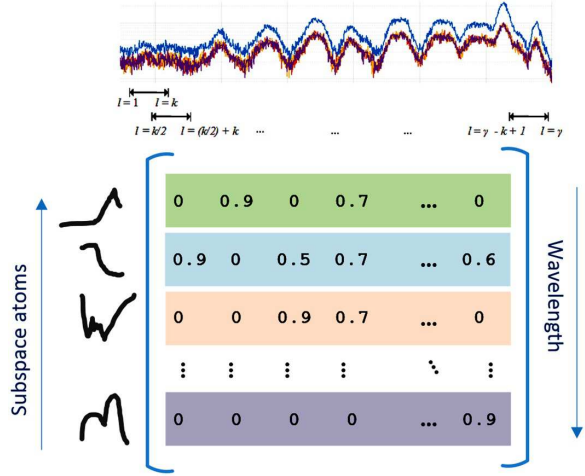}
  \caption{ Conceptual diagram of multi-wavelength \textit{atom-layer} relationship matrix (coefficient matrix) which is learnt for several over-lapping sub-sequences (columns) on range-amplitude ladar signals.} 
  \label{fig:semantic_learning}
 \end{figure*}

The algorithm in its entirety is illustrated above. Repeat step 2 for each individual layer (sub-sequence) and stack the coefficient matrix $C_{\lambda, i}$, where $\lambda = \left[1,2,...,\Lambda\right]$(output of Algorithm~\ref{alg:sad}). So, for layer 1 $F_{\Lambda,1}$ matrix can be expressed as: 
\begin{equation} \label{F_matrix}
F_{\Lambda,1}=	\begin{bmatrix}C_{\lambda, 1} \\
\vdots\\
\vdots \\
C_{\Lambda, 1}
\end{bmatrix}\in \mathbb{R}^{\left(\Lambda\times \rho\right) \times \eta},
\end{equation} 
and  a set of such matrices can be represented as: 
\begin{equation} \label{eq:W_to_F_final}
\mathbf{F}_{train}=\left\{ \underset{W_{\lambda,1}}{\underbrace{F_{\lambda,1}}}\,\,,\,\,\underset{W_{\lambda,2}}{\underbrace{F_{\lambda,2}}}\,\,,\,\,...\,\,,\,\,...\,\,,\underset{W_{\Lambda, B}}{\underbrace{F_{\Lambda,B}}}\right\}. 
\end{equation}
The label under-set each $F_{\Lambda,B}$ shows which layer it belongs to. 
Conceptual diagram showing how $\mathbf{F}_{train}$ matrix is constructed is shown in Figure \ref{fig:semantic_learning}. The final step is to reduce the dimensionality in the $\Lambda$ dimension and learn the relationships across different layers at the same time. This is shown below. 
\subsection{Step 3: Relationship Learning \& Anomaly Detection} \label{sec:rln}
The Singular Value Decomposition (SVD) is used to decompose $F_{train}$: 
\begin{equation} \label{eq:orig_svd}
	\mathbf{F}_{train} = \mathit{U}\Sigma\mathit{V}^{T}, 
\end{equation}
where, $\mathit{U} = \left[\mathbf{u_{1}},...,\mathbf{u_{\Gamma \times \rho}}\right]$ ($\rho=$ number of training examples) and $\mathit{V}=\left[\mathbf{v_{1}},...,\mathbf{v_{\eta}}\right]$ ($\eta=$ number of basis vectors) are sets of singular vectors, and $\Sigma$, a diagonal matrix of the singular values. In this representation, the vectors of $\mathit{U}$ correspond to the elementary atoms that make up a ladar signal and vectors of $\mathit{V}$ corresponds to a tree/forest layer (window) in the semantic space. To preserve only $\mathit{q}$ essential semantic relationships, one needs to approximate the \textit{atom-layer} relationship matrix as follows: 
\begin{equation} \label{eq:svd_approx}
	\mathbf{F}_{train} \approx \mathit{U_{q}}\Sigma_{q}\mathit{V_{q}^{T}}. 
\end{equation}

In the training phase apply (\ref{eq:svd_approx}) to the atom-layer (coefficient) matrix $\mathbf{F}_{train}$ resulting in $\mathit{U}_{q}$ and $\Sigma_{q}$. During testing, the atom-layer coefficient matrix $\mathbf{F}_{test}$ is transformed in to a lower-dimensional semantic space using: 
\begin{equation} \label{eq:svd_linear_trans}
	\mathbf{\hat{F}}_{test} = \Sigma_{q}^{-1}\mathit{U}_{q}^{T}\mathbf{F}_{train},
\end{equation}
and reconstructed in to the original space using: 
\begin{equation}	 \label{eq:svd_reconstruct}
	\mathbf{\tilde{F}} = \mathit{U}_{q}\Sigma_{q}\mathbf{\hat{F}}_{test}.
\end{equation}
This step completes the mapping of $\mathbf{W}\to\mathbf{F}$ which results in a mapping $\mathbf{F}_{train} \in q \times \eta $ where, $q \ll \left(\Lambda \times \rho\right)$. 

\paragraph{Anomaly Detection and Score} It should be noted that while the low-rank nature of $\mathit{U}_{q}$ and $\Sigma_{q}$ captures only the essential semantic relationships,  (\ref{eq:svd_reconstruct}) cannot reconstruct the original atom-layer coefficient matrix perfectly. The underlying intuition here is that if the co-occurrence relationships in the test ladar samples are different to those in the training data set, the reconstruction error will be large, i.e., the anomalous elements of the coefficient matrix will have significant large value. This is computed by calculating the distance between the learnt representation and reconstructed signal. The reconstruction error on each individual layer is computed: 
\begin{equation} \label{eq:anomaly_score}
	\hat{A}_{score} = ||\mathbf{F}_{train} - \mathbf{\tilde{F}}||_{2}^{2},
\end{equation}
Equation (\ref{eq:anomaly_score}) computes a score with regard to an individual layer (sliding window), an anomaly score for each FW-MSL signal with $\mathit{N}$ layers can be written as: 
\begin{equation} \label{eq:anomaly_avg_score}
	A_{score} = \frac{1}{N}\sum_{1}^{N}\tilde{A}_{score}
\end{equation}
\paragraph{Note} If the classification mode is selected, the above steps are followed except that training labels for different classes are made available. In such a scenario, the discriminative objective function (\ref{eq:fisher_objective}) is used to learn atom-layer relationships and assign them class labels. For each test sample a K-NN match is computed in order to predict its class label and then compute (\ref{eq:svd_linear_trans} and \ref{eq:svd_reconstruct}). Initial experiments show that this reduces false alarms but at an expense of CPU time in solving (\ref{eq:fisher_objective}). Due to space constraints, this paper only highlights only anomaly detection scores. 

\begin{figure}[!t] \label{fig:sim}
\centering
 \includegraphics[width=17.2cm, height=4.5cm]{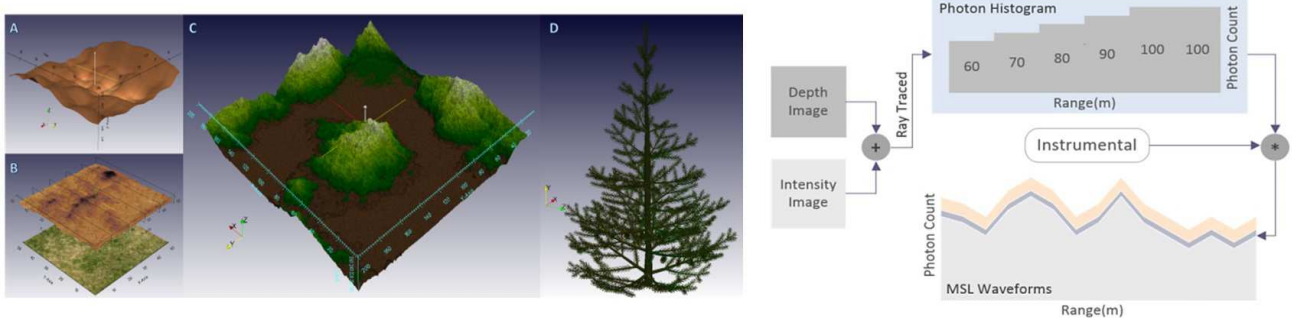}
  \caption{(Left) Natural terrain generation method using fractal based Brownian Motion Fields. In (A) A smooth dessert terrain with a high Hurst index $H$. The two-step terrain generator process is shown in (B), layer 1 is a factorial Brownian filed (fBF) and layer 0 is a texture layer that is mapped to the elevation model. (C) A combination of several two-step processes (shown in (B)) in creating a combinatorial terrain with elevated hills and rough terrain. (D) Output of a closed-source tree modeller, Onyxtree that is used in this study for several conifer species. (Right) Once the scene is rendered, it is show how the depth and intensity information can be used to generate synthetic MSL waveforms.} 
  \label{fig:data_gen}
 \end{figure}
\section{\uppercase{Synthetic Data and Experiments}} \label{sec:Experiments}
When scanning an urban setting or dense forest we employ two scanning schemes (See Figure  \ref{fig:scan_schemes}); firstly, a \textit{push-broom} style scan where we carry out a dense grid style scan and secondly, a \textit{flash} style scan with larger beam resulting in a large foot-print scan that illuminates an entire tree in one go. In Figure \ref{fig:data_gen} we illustrate the waveform generation process based on depth image and intensity image. A synthetic scene is ray-traced using a custom modified Povray\footnote{http://www.pov-ray.org} that also generates a depth image. Equation (\ref{eq:ladar_power}) is used to convolve the scene with a Gaussian instrumental (3-5ns) to generate synthetic FW-MSL waveforms. For the sake of simplicity the same instrumental is used across all wavelengths. The received power $P_{a}\left(\gamma\right)$ of a Ladar signal after interacting with $S$ surfaces, at a range $R$, with $D_{a}$ as the aperture diameter of the receiver sensor can be expressed as: 
\begin{equation} \label{eq:ladar_power}
	P_{a}\left(\gamma\right) = \sum_{i=1}^S \frac{D_{a}^2}{4\pi\lambda^{2}R_{i}^{4}}\, P_{d}\left(\gamma\right)\, \ast \Gamma\left(\gamma\right) \, \ast E\left(\gamma\right)\,\ast\sigma_{i}\left(\gamma\right), 
\end{equation} 
where $\ast$ denotes a convolution operation over $P_{d}\left(\gamma\right)$, the transmitted power, $\Gamma\left(\gamma\right)$, the system impulse, $E\left(\gamma\right)$ , environmental aerosol contribution and $\sigma_{i}\left(\gamma\right)$, is the backscattering cross-section.
\subsection{Synthetic Data - Multi-Beam Modeller} \label{sec:Data}
The temporal shape of the received signal often referred to as a waveform can be written as a sum of all the echoes from $S$ surfaces. Some of the resulting waveforms for a single-wavelength are shown in Figure \ref{fig:example} with range as its x-axis and amplitude (photon count) on the y-axis. Similar waveforms are generated for 4, 8, 16 and 32 wavelengths. 

In this section we first present a method to model 3D forest-terrain and then generate realistic trees using a closed-source popular tree modeller, OnyxTree\footnote{http://www.onyxtree.com}. Two sets of datasets are collected, one which contains up to 3 different conifer tree species (Picea Orientalis, Pseudo Suga M2 and Picea Engelmannii) with variable height and Leaf Area Index (LAI); second, we populate the same environment with man-made objects such as vehicles at random locations. In order to make the setting more realistic we add random undergrowth in between the trees. The reflectance spectra of all the materials used in the experiments (\textit{man-made}\footnote{The Aster2.0\cite{baldridge2009aster} spectral library provided by JPL, NASA is used for reflectance values of man-made objects.}: brick, granite, steel, roof-tiles and \textit{natural objects}\footnote{Passive measurements of tree samples under laboratory conditions\cite{wallace2014} was carried out using a stabilized light source and spectrometer, with several needles and bark samples.}: conifer bark, needle and Aridisol (soil)) is illustrated in Figure \ref{fig:material_spectra}. Next, a brief note on how natural forest terrains are generated is presented. 
 \begin{figure}[!t]
\centering
 \includegraphics[width=17cm, height=5.2cm]{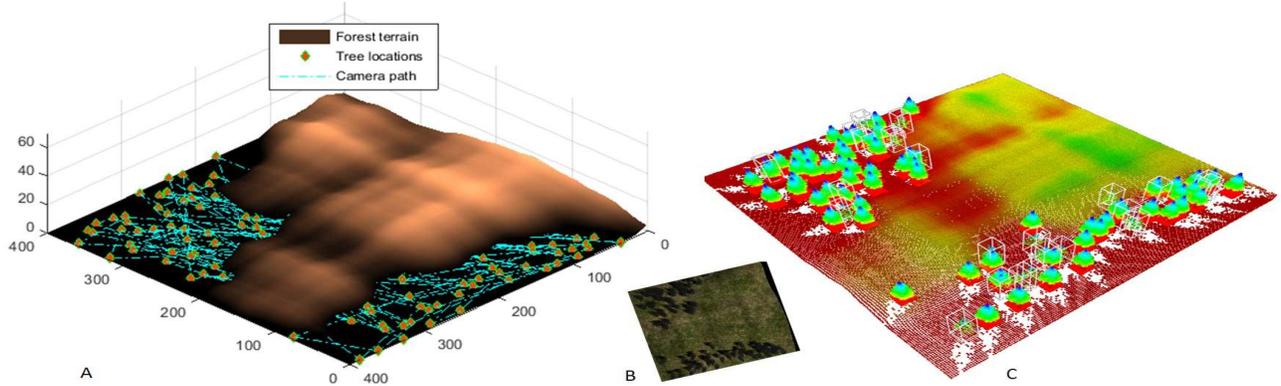}
  \caption{Rendered scene with conifer species. A) Terrain generated along with tree locations and camera path. B) Rendered scene (2D Image). C) 3D point cloud with elevation shaded (light to dark).} 
  \label{fig:rendered_scene}
 \end{figure}
 \begin{figure}[!h]
\centering
 \includegraphics[width=12cm, height=5cm]{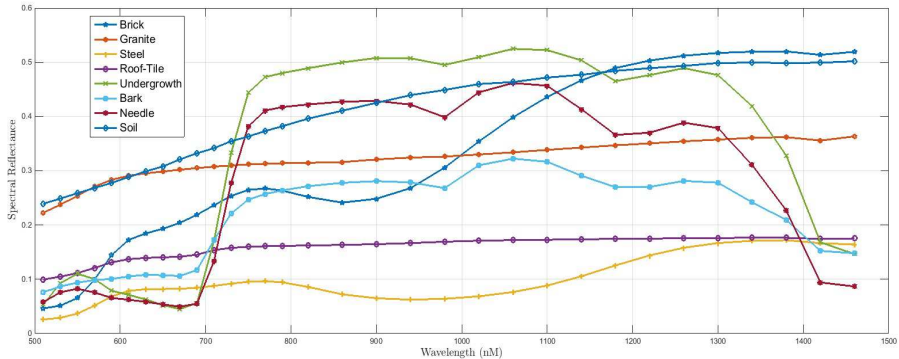}
  \caption{Measured material spectral reflectance for several \textit{natural}: tree bark, needle and soil samples along with data from the Aster spectral library for selected \textit{man-made objects}: building material and galvanized steel.} 
  \label{fig:material_spectra}
 \end{figure}
\paragraph{Natural Terrain Modeller} In nature we find natural objects tend to follow a symmetry pattern and self-similarity. Mandelbrot \cite{mandelbrot1983fractal} in his influential paper linked such symmetry to a mathematical phenomenon called fractals. Following this method, natural terrains can be modelled using a fractal interface. In this work a forest terrain as a simple continuous-time stochastic process and model it as a fractal or factorial Brownian Field (fBF). The terrains are made up of two layers. Layer one, is an elevation field and texture is mapped to the elevation model in layer two. Figure \ref{fig:data_gen} shows examples of different terrains generated. An elevation function is defined using a generalised Brownian motion without independent increments: 
\begin{equation}\label{eq:fBF}
	E \left[ B_{t_{1}},B_{t_{2}} \right] = \left\langle B_{t_{1}}B_{t_{2}}\right\rangle = \frac{\sigma^{2}}{2} \left[ t^{2H} - (t_{2} - t_{1})^{2H} + s^{2H}\right]
\end{equation} where $B_{t}$ is a continuous-time Gaussian Process on $[0,\mathit{T}]$ and $\sigma$, 
The roughness parameter $\mathit{H}$, called Hurst index is varied in order to control the raggedness of the motion and this can result in a variety of surfaces such as smooth dessert type terrains and mountainous regions. Combinations of more than one surfaces are used to build rocky terrains and also by controlling the shape and geometry of the motion by defining a random moment matrix and an asymmetry property to the matrix. The steps involved are illustrated in Figure \ref{fig:sim}. The terrain grid along with elevation values are generated in MATLAB. Different tree species, man-made objects undergrowth along with generated terrain are then rendered using modified Pov-ray.
 
\paragraph{Man-made Objects} \label{sec:build_gen}
Most of the man-made objects come from manufacturing processes or architectural constructions and are a complex arrangement of basic ensembles (cylinders, spheres or planes). Such objects are used in the experiments in order to study the spectral responses of different surfaces (elevated areas, small objects and vegetation surfaces) with multi-beam and multi-spectral Ladar sensor. Man-made structures with elevated roofs and planar platforms along with vehicles are rendered and placed in a forest setting.
\begin{figure}[!t]
\centering
 \includegraphics[width=17cm, height=11cm]{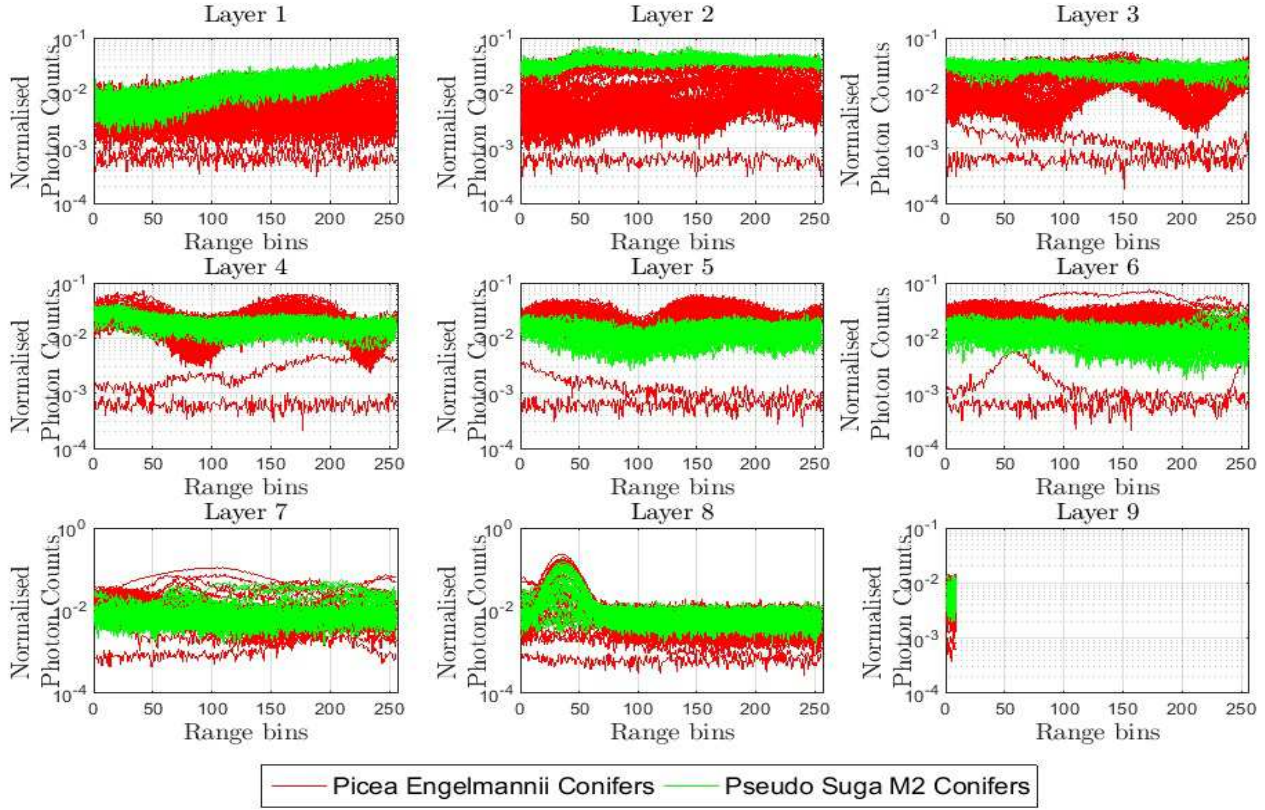}
  \caption{Test FW ladar signals belonging to 2 different conifer species for a single wavelength partitioned in to 9 layers using $\mathbf{L} \to \mathbf{W}$. In one of the experiements 105 sample belonging to two different conifer species are used. Anomalies are introduced in the form of man-made objects at different depths and layers.} 
  \label{fig:rec_error}
 \end{figure}

\subsection{Results on Synthetic Data}
Scenes rendered using the proposed modeller, as illustrated in Figure \ref{fig:rendered_scene} are used for experimentation. This small plot of forest is made up of two conifer tree species and several man-made objects are placed in between. These objects are vehicles closer to the ground and camouflaged with a random undergrowth introduced in between the trees. 

An overhead surveillance flight simulation is assumed. Flight path is traced and several regions of the forest are illuminated using large foot-print waveforms and backscattered returns are recorded using \ref{eq:ladar_power}. Out of 105 test samples 23 FW-MSL signals contain several anomalies. The detections are shown in Figure \ref{fig:rec_error}. 

Further analysis is carried out only on the detected anomalous waveforms in order to produce a dense 3D point cloud. Figure \ref{fig:anom_3D} illustrates some of these detections. Also,  shown is an anomalous waveform and one of the learnt models which is constructed using the learnt dictionaries for each layer. The anomalies, both in the point cloud and waveform have been highlighted with a box. The conifer in Figure \ref{fig:anom_3D}(A) conceals a large part of a vehicle that has a two surfaces at varying depths. This is obvious in \ref{fig:anom_3D}(D), highlighted with a box between layer 7 and 8. Subtle anomalies are also detected, e.g., in Figure \ref{fig:anom_3D}(C) where only a tiny corner of the vehicle falls within the sensor field of view. 
\begin{figure}[!t]
\centering
 \includegraphics[width=17cm, height=12cm]{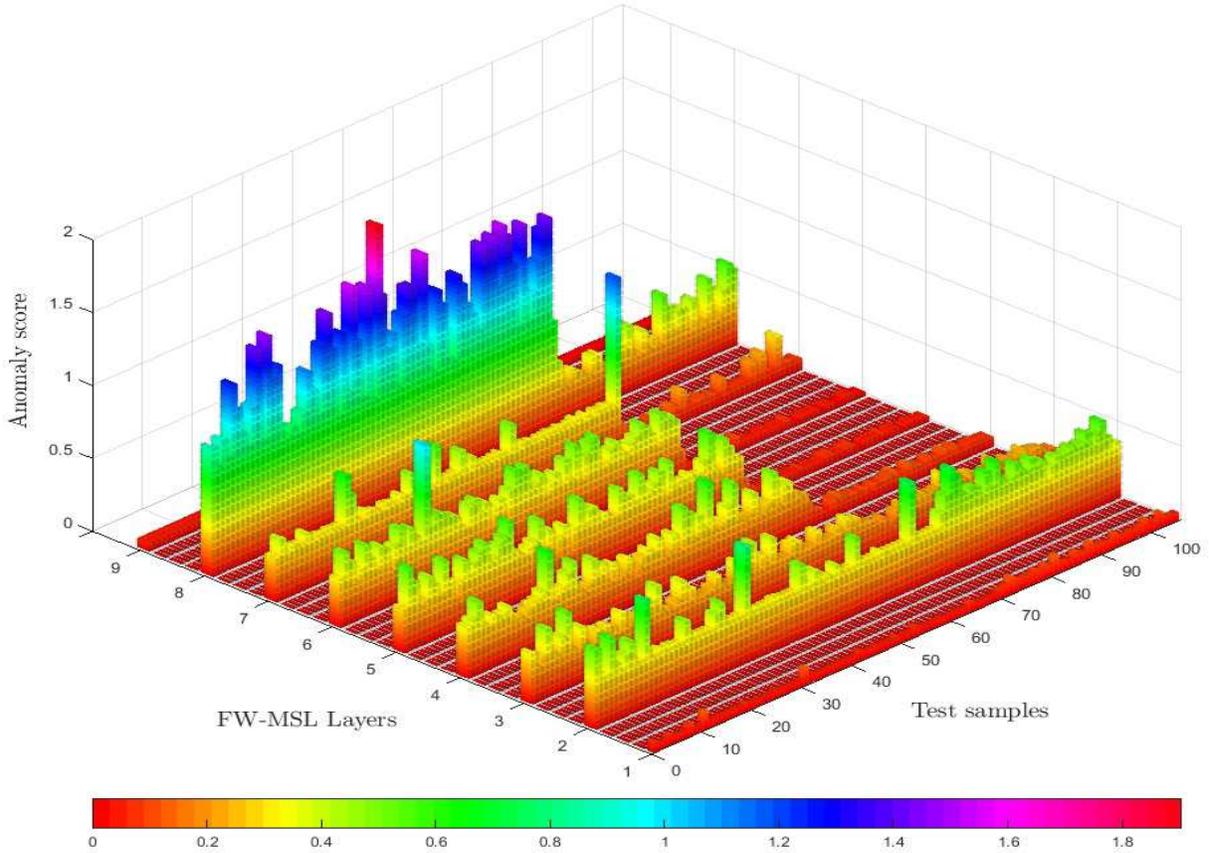}
  \caption{Layer wise reconstruction error computed using \ref{eq:svd_reconstruct} is illustrated here. These results also label the species of the trees using a discriminative optimisation routine \ref{eq:fisher_dis}. Spectral and structural anomalies are added in the form of man-made objects and bare ground. Most of these objects are closer to the ground, Layer 7 and 8 clearly show the existence of these anomalies at 15, 20, 34, 69, 103, etc. Samples 34 and 69 record high reconstruction error as this is bare forest floor with some undergrowth. Layer 1 and 9 record the lowest reconstruction error as these layers record  mostly background and system noise.} 
  \label{fig:rec_error}
 \end{figure}
\section{\uppercase{Conclusion \& Future Work}} \label{sec:Conclusion}
\begin{figure}[!t]
\centering
 \includegraphics[width=17cm, height=11cm]{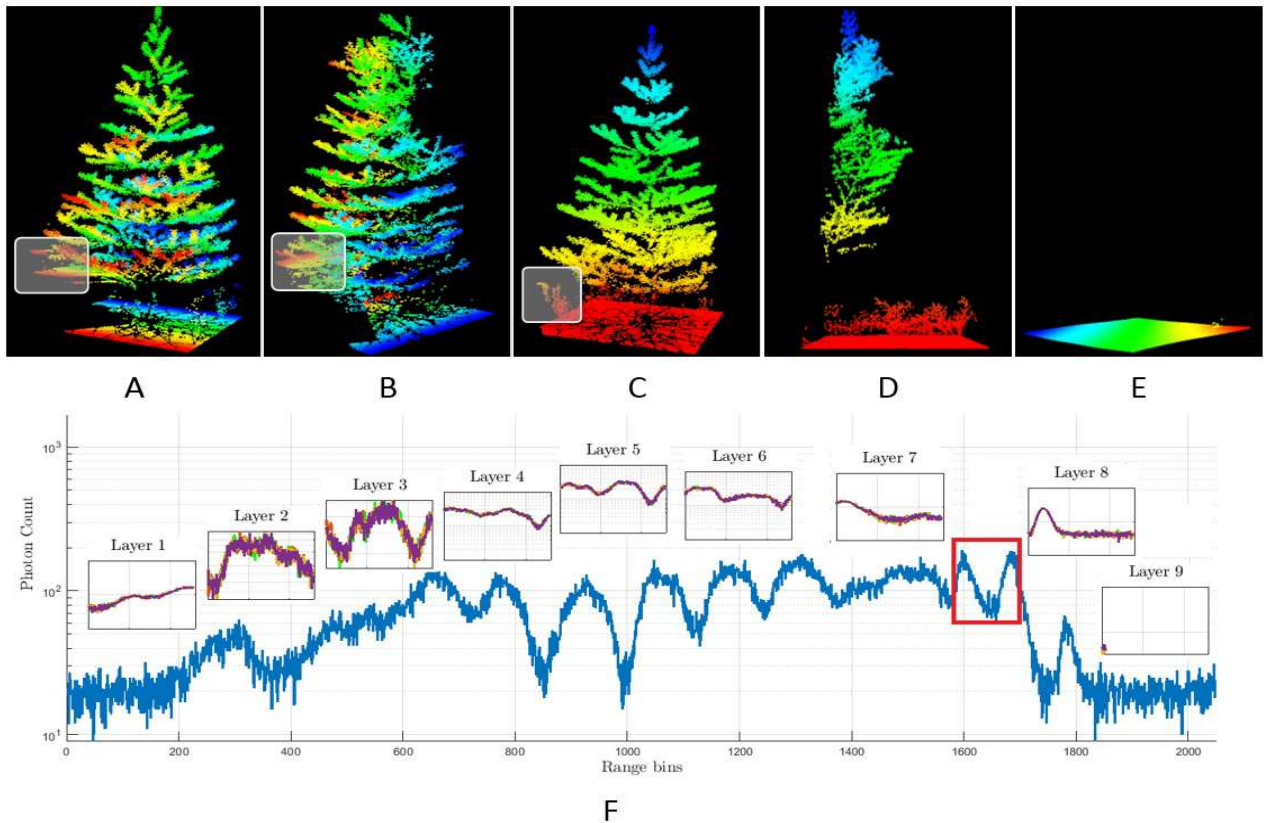}
  \caption{(A) A 3D point cloud of test sample 37, which is detected as an anomaly using the proposed algorithm confirms the presence of large two surfaced object. The detection peaks are shown in between layer 7 and 8. (B) Large section of a man-made object detected on test sample 41. (C) A small subtle anomaly is being detected on layer 8 of test sample 21. (D) and (E) are two anomalies where the ground spectra along with undergrowth dominates the coefficient space. (F) Layer -wise reconstruction of representative models are shown along with test sample 37. The reconstruction is build using dictionaries learnt on individual layers for 32 different wavelengths.} 
  \label{fig:anom_3D}
 \end{figure} 
 
This paper suggests a two phase approach to anomaly detection and target detection in dense cluttered environments. This work is the first piece of work, that the authors are aware of which aims to look for spectral and depth anomalies in raw FW-MSL waveforms. 8 different spectra are introduced in a cluttered forest scenes in the form of man-made objects and vegetation. The suggested approach is widely applicable no matter the sensor modality. This has been considered as a precursor to find \textit{interesting} signals, worthy of more detailed sampling and analysis in order to detect man-made objects, e.g. vehicles under foliage, or mines underwater. 

The proposed approach is not restrictive and separate passive hype-spectral imagery registered with Ladar measurements can also be used as long as the training and test measurements are registered or geo-referenced with respect to the sensor. Any number of tree species can be learnt within this framework and as a by-product, the proposed approach allows segmentation of individual foot-prints for semantic scene labelling. This will be explored in detail in future works.

FW-MSL data generation method proposed in this paper generates synthetic data using natural terrain modeller along with a ray-tracer. Comparing this to past MSL work on studying tree physiology \cite{Wallace2014}, the synthetic data is as good as real measurements. Future work will concentrate on testing the proposed algorithm on real FW-MSL data where the authors aim to emulate an aerial surveillance scenario with real vehicles and building material concealed under dense foliage. Additional work looking into robust relationship learning algorithms in order to limit the number of false detections could be highly beneficial.  

\appendix    
\acknowledgments     
 
This work was supported by the Engineering and Physical Sciences Research Council (EPSRC) Platform Grant EP/J015180/1 and the MOD University Defence Research Collaboration in Signal Processing Grant EP/K014277/1.


\bibliography{article}   

\begin{thebibliography}{10}

\bibitem{reitberger2006}
J.~Reitberger, P.~Krzystek, and U.~Stilla, ``Analysis of full waveform lidar
  data for tree species classification,'' {\em International Archives of
  Photogrammetry, Remote Sensing and Spatial Information Sciences}~{\bf
  36}(Part 3), pp.~228--233, 2006.

\bibitem{brodu2012}
N.~Brodu and D.~Lague, ``{3D} terrestrial lidar data classification of complex
  natural scenes using a multi-scale dimensionality criterion: Applications in
  geomorphology,'' {\em ISPRS Journal of Photogrammetry and Remote
  Sensing}~{\bf 68}, pp.~121--134, 2012.

\bibitem{jutzi2006}
B.~Jutzi and U.~Stilla, ``Range determination with waveform recording laser
  systems using a {Wiener} filter,'' {\em Journal of Photogrammetry and Remote
  Sensing}~{\bf 61}(2), pp.~95--107, 2006.

\bibitem{mallet2011}
C.~Mallet, F.~Bretar, M.~Roux, U.~Soergel, and C.~Heipke, ``Relevance
  assessment of full-waveform lidar data for urban area classification,'' {\em
  ISPRS Journal of Photogrammetry and Remote Sensing}~{\bf 66}(6),
  pp.~S71--S84, 2011.

\bibitem{chehata2009}
N.~Chehata, L.~Guo, and C.~Mallet, ``Contribution of airborne full-waveform
  lidar and image data for urban scene classification,'' in {\em
  16\textsuperscript{th} IEEE International Conference on Image Processing
  (ICIP), 2009},  pp.~1669--1672, IEEE, 2009.

\bibitem{wallace2010}
A.~M. Wallace, J.~Ye, N.~J. Krichel, A.~McCarthy, R.~J. Collins, and G.~S.
  Buller, ``Full wave form analysis for long-range {3D} imaging laser radar,''
  {\em EURASIP Journal on Advances in Signal Processing}~{\bf 2010}, p.~33,
  2010.

\bibitem{wallace2014}
A.~M. Wallace, A.~McCarthy, C.~J. Nichol, X.~Ren, S.~Morak,
  D.~Martinez-Ramirez, I.~H. Woodhouse, and G.~S. Buller, ``Design and
  evaluation of multispectral lidar for the recovery of arboreal parameters,''
  2014.

\bibitem{harsanyi1994hyperspectral}
J.~C. Harsanyi and C.-I. Chang, ``Hyperspectral image classification and
  dimensionality reduction: an orthogonal subspace projection approach,'' {\em
  Geoscience and Remote Sensing, IEEE Transactions on}~{\bf 32}(4),
  pp.~779--785, 1994.

\bibitem{candes2008introduction}
E.~J. Cand{\`e}s and M.~B. Wakin, ``An introduction to compressive sampling,''
  {\em Signal Processing Magazine, IEEE}~{\bf 25}(2), pp.~21--30, 2008.

\bibitem{scholkopft1999fisher}
S.~Mika, G.~R{\"a}tsch, J.~Weston, B.~Sch{\"o}lkopf, K.~M{\"u}ller, Y.-H. Hu,
  J.~Larsen, E.~Wilson, and S.~Douglas, ``Fisher discriminant analysis with
  kernels.,'' in {\em Neural Networks for Signal Processing},  pp.~41--48,
  1999.

\bibitem{tropp2006algorithms}
J.~A. Tropp, A.~C. Gilbert, and M.~J. Strauss, ``Algorithms for simultaneous
  sparse approximation. part i: Greedy pursuit,'' {\em Signal Processing}~{\bf
  86}(3), pp.~572--588, 2006.

\bibitem{pati1993orthogonal}
Y.~C. Pati, R.~Rezaiifar, and P.~Krishnaprasad, ``Orthogonal matching pursuit:
  Recursive function approximation with applications to wavelet
  decomposition,'' in {\em Signals, Systems and Computers, 1993. 1993
  Conference Record of The Twenty-Seventh Asilomar Conference on},  pp.~40--44,
  IEEE, 1993.

\bibitem{baldridge2009aster}
A.~Baldridge, S.~Hook, C.~Grove, and G.~Rivera, ``The aster spectral library
  version 2.0,'' {\em Remote Sensing of Environment}~{\bf 113}(4),
  pp.~711--715, 2009.

\bibitem{mandelbrot1983fractal}
B.~B. Mandelbrot, ``Fractal aspects of the iteration of z→ $\lambda$z (1-z)
  for complex $\lambda$ and z,'' {\em Annals of the New York Academy of
  Sciences}~{\bf 357}(1), pp.~249--259, 1980.

\end{thebibliography}
\bibliographystyle{spiebib}   

\end{document}